\documentclass[aps,prd,preprintnumbers,superscriptaddress,nofootinbib,floatfix,twocolumn,notitlepage]{revtex4-2}
\usepackage{graphicx}  
\usepackage{dcolumn}   
\usepackage[table]{xcolor} 
\usepackage{makecell}   
\usepackage{bm}        
\usepackage{epsfig,amsmath,amssymb,verbatim,mathrsfs,array,layout,textcomp,amssymb,latexsym,slashed}
\usepackage{xcolor}
\usepackage[colorlinks=true,citecolor=blue,urlcolor=blue,linktocpage=true,
linkcolor=blue]{hyperref}
\usepackage{cleveref} 
\usepackage[utf8]{inputenc}
\usepackage{multirow}
\usepackage{xspace}
\usepackage{booktabs} 
\usepackage{siunitx} 
\usepackage{tabularx} 
\usepackage{physics} 
\usepackage{dsfont} 
\usepackage{physunits} 

\usepackage[inkscapeformat=png]{svg}
\usepackage[normalem]{ulem} 

\newcommand{\cL}{\mathcal{L}}

\newcommand{\BR}{\mathrm{BR}}

\newcommand{\eg}{e.g.} 

\newcommand{\GeV}{\textrm{GeV}}
\newcommand{\TeV}{\textrm{TeV}}

\newcommand{\be}{\begin{equation}}
\newcommand{\ee}{\end{equation}}
\newcommand{\bea}{\begin{eqnarray}}
\newcommand{\eea}{\end{eqnarray}}

\newcommand{\mysec}[1]{\textbf{#1}}

\definecolor{Ecolor}{HTML}{008000}
\definecolor{Bcolor}{HTML}{c8ab37}
\definecolor{edm}{HTML}{c83771}

\definecolor{onlyP}{HTML}{003f5c}
\definecolor{onlyN}{HTML}{bc5090}
\definecolor{bothPN}{HTML}{ffa600}

\begin{document}

\title{Probing new hadronic forces with heavy exotic atoms}
 
\author{Hongkai Liu}
\email{hliu6@bnl.gov}
\affiliation{Physics Department, Technion -- Israel Institute of Technology, Haifa 3200003, Israel}
\affiliation{High Energy Theory Group, Physics Department, Brookhaven National Laboratory, Upton, NY 11973, USA}

\author{Ben Ohayon}
\email{bohayon@technion.ac.il}
\affiliation{Physics Department, Technion -- Israel Institute of Technology, Haifa 3200003, Israel}

\author{Omer Shtaif}
\email{omer.shtaif@campus.technion.ac.il}
\affiliation{Physics Department, Technion -- Israel Institute of Technology, Haifa 3200003, Israel}

\author{Yotam Soreq}
\email{soreqy@physics.technion.ac.il}
\affiliation{Physics Department, Technion -- Israel Institute of Technology, Haifa 3200003, Israel}

\begin{abstract}
We explore the potential of precision spectroscopy of heavy exotic atoms where electrons are substituted by negative hadrons to detect new force carriers with hadronic couplings.
The selected transitions are unaffected by nuclear contact terms, thus enabling highly accurate calculations using bound-state QED, provided that the nuclear polarization is under control.
Alternatively, we demonstrate that the dipole polarizability, a fundamental property of nuclei, can be extracted from the spectroscopy of exotic atoms in a novel way by combining two transitions while maintaining high sensitivity to new physics.
Based on existing data, we extracted world-leading bounds on mediator masses ranging from  $0.1\,\MeV$ to $10\,\MeV$ for two benchmark models and show that forthcoming experiments could enhance the sensitivity to new physics by two orders of magnitude.
\end{abstract}

\maketitle

\mysec{Introduction}.
High-precision atomic spectroscopy plays a key role in fundamental physics. 
Two applications are determining fundamental constants~\cite{2005-Const,2024-CODATA} and searching for physics beyond the standard model~(BSM), \eg~\cite{2010-SI, 2010-PSAS, Safronova:2017xyt, Cong:2024qly}.
Such new physics~(NP) is well motivated by experimental evidence and strong theoretical arguments and can be associated with low-mass states well below the GeV scale; see \eg~\cite{EuropeanStrategyforParticlePhysicsPreparatoryGroup:2019qin}. 
NP searches in atomic spectra can be broadly categorized into two types. 
The first category involves scenarios where (approximated) symmetries and/or factorization properties significantly amplify the NP signal compared to the standard model~(SM) (see \eg~\cite{Wood:1997zq,2000-Var,Bouchiat:2004sp,Arvanitaki:2014faa,Delaunay:2016brc,Berengut:2017zuo,Baruch:2024fbh,Roussy:2022cmp}). 
The second category involves identifying the NP signal by a meticulous comparison between the SM predictions and measurements as in~\cite{2010-SI,2010-PSAS, 2011-ALPs, Delaunay:2022grr}.
This work falls into the second category.

We consider the effect of a new spin-independent interaction coupled to hadrons on the spectra of exotic atoms, where the electrons are replaced by a single $\bar{p}$, $\pi^-$, or a $K^-$, and derive novel bounds and projections for future measurements. 
These atomic systems present several advantages over electronic or muonic atoms:
(i)~they allow the study of purely hadronic interactions, unaffected by electronic and muonic couplings;
(ii)~they enable probing of shorter distances than those in molecules or electronic atoms, enhancing the sensitivity to the detection of new force mediators with higher masses, $\sim10\,\MeV$ compared to $\sim4\,\keV$\footnote{We use natural units: $\hbar=c=1.$}; 
(iii)~they facilitate straightforward theoretical comparisons, as electrons are absent or their influence is greatly reduced~\cite{Simons_1988}, making these simple and highly charged systems amenable to precise QED calculations~\cite{1982-1ppm, Paul:2020cnx,Zatorski:2022sze, SANTOS2005206, 2005-QDEK}.
However, these searches face theoretical and experimental challenges. 

The transition energies are proportional to $\alpha^2 Z^2 m_H$ ($Z$ is the atomic number, $\alpha$ the fine-structure constant, and $m_H$ the hadron mass), placing them primarily in the x-ray range.
Measuring these energies with high accuracy is challenging due to the low resolving power of solid-state detectors~\cite{2008-Noble}, and the narrow band and low efficiency of high-resolution wavelength measurement techniques~\cite{2019-DCS}.
These difficulties are compounded with the limited rates at which exotic particle beams are available.
Nevertheless, high-resolution measurements of the x-ray energies of exotic atoms have been made using crystal spectrometers~\cite{1982-Ruckstuhl, WEBER1981343, WEBER1982361, 1983-Quad, 1984-Ruckstuhl, BEER1985365, 1985-13C, 1985-StrongPi, BELTRAMI1986679, 1985-7Li, leisi1986muonic, gilot1987crystal, GOTTA1993645, 1999-StronPBAR, GOTTA2004133, hirtl2021redetermination, ANAGNOSTOPOULOS1999c305, 2003-PbarNe, 2000-Expl, 1998-StrongPi, LENZ199850, Trassinelli:2016kki, JECKELMANN1986709, gotta2014precision}.
Moreover, precision x-ray spectroscopy of exotic atoms is currently undergoing a paradigm shift due to the introduction of cryogenic microcalorimeter detectors~\cite{2016-HEATES, 2022-KHe, 2024-uNe, 2024-QUARTET, 2024-Ar, 2025-TES, 2025-PAX}, which offer high quantum efficiency and exceptional resolution~\cite{2024-TES, 2024-Unger}. 

On the theory side, it has been difficult to extend NP searches with atomic spectroscopy to high NP mediator masses. 
Here, factors such as nuclear structure (size, shape, deformation, etc.) and, in certain systems, strong interaction effects~\cite{1985-StrongPi, 1997-Strong, 1999-StronPBAR, AUGSBURGER1999149, ANAGNOSTOPOULOS1999c305, 1998-StrongPi, hirtl2021redetermination, 2022-KHe}, become significant while being poorly estimated due to the non-perturbative nature of QCD at low energies.
However, circular states, that is, states of a high principal quantum number $n$, and a maximal angular momentum $l=n-1$ are not sensitive to the short-distance nuclear effects~\cite{1997-Strong, Paul:2020cnx}, thus they are amenable to state-of-the-art QED calculations, \eg~\cite{Paul:2020cnx}.

For these states, the least known nuclear effect is the long-range part of the nuclear polarization~(NPol)~\cite{Ericson:1972nhh, Zatorski:2022sze}, which depends on a fundamental nuclear property: static electric dipole polarizability, $\alpha_{E}^N$.
As a single measurement cannot distinguish NPol from NP, we demonstrate how the use of two transitions enables extracting both with minimal loss in sensitivity.
Measurements of $\alpha_{E}^N$ are interesting on their own for nuclear structure studies~\cite{2017-alphad}, as well as for determining the contribution of the SM to nonlinearities of King plots~\cite{2022-NPKP}, which limits the sensitivity to new ``fifth forces"~\cite{2024-CaKP}.
Simultaneous measurements of two transitions also benefit from the broadband capabilities of microcalorimeter detectors.

Below, we extract new bounds on the BSM hadronic interactions from existing data in $\bar{p}\,$Ne~\cite{1999-StronPBAR}, $\bar{p}\,$Pb~\cite{1975-PbExp,Borie:1983nlf} and $\pi^-\,^{14}$N~\cite{Trassinelli:2016kki} and derive projections to $\bar{p}\,^{20}\,$Ne, $\bar{p}\,^{132}\,$Xe and $K^-\,^{20}\,$Ne which are targets of ongoing measurements~\cite{2019-Kaon,2024-Prospects,2024-KNe,2024-169,2024-PAX,2025-PAX}. 
In particular, we show that these can be the leading probes of certain models, where the bounds from rare kaon decay are insignificant~\cite{Delaunay:2025lhl}.
Although there are bounds existing from hadronic atom spectroscopy, \eg{} $\bar{p}\,^4$ He~\cite{ASACUSA:2016xeq,Germann:2021koc},
we analyze atoms that probe higher mediator masses ($\gtrsim100$~keV) of the new boson, corresponding to a shorter interaction range. 
\begin{table*}[t]
    \centering
    \begin{tabular}{l c c c c c  c}
    \hline\hline
    Bound-state      & $(n_{i},l_{i})$ & $(n_{f},l_{f})$ & $E^{\rm{SM,LO}}_{n,n-1}\,$[keV] & $\Delta_{n,n-1}/E^{\rm{SM,LO}}_{n,n-1}\,$[ppm] & $1/r_{n_f}\,$[MeV] & $ E^{\rm NPol}_{n,n-1}/E^{\rm SM,LO}_{n,n-1}$[ppm]\\\hline  
    $\pi^-~^{14}$N   & (5,4)   & (4,3)   & 4.05                & $3.9\pm1.7$\,\cite{Trassinelli:2016kki}             &0.44  & $0.005$    \\\hline
    $K^-~^{20}$Ne    & (6,5)   & (5,4)   & 15.6                & $\pm1\,$                                            &1.4   & $<0.5$     \\\hline
    $\bar p~^4$He    & (32,31) & (31,30) & 5.11$\times10^{-3}$ & $(2.2\pm2.3)\times10^{-3}\,$~\cite{ASACUSA:2016xeq} &0.011 & $10^{-8}$  \\
    $\bar p~^{20}$Ne & (13,12) & (12,11) & 2.44                & $14\pm23\,$\cite{1999-StronPBAR}                    &0.45  & $0.005$    \\
    $\bar p~$Pb      & (12,11) & (11,10) & 221                 & $109\pm77\,$~\cite{Borie:1983nlf,1975-PbExp}        &4.6   & $20$       \\
    $\bar p~$Pb      & (11,10) & (10,9)  & 290                 & $114\pm72\,$~\cite{Borie:1983nlf,1975-PbExp}        &5.5    & $40$       \\\hline
    $\bar p~^{20}$Ne & (6,5)   & (5,4)   & 29.0                & $\pm1\,$                                            &2.6   & $1$        \\
    $\bar p~^{132}$Xe& (11,10) & (10,9)  & 125                 & $\pm1\,$                                            &3.7   & $8$        \\
    $\bar p~^{132}$Xe& (10,9)  & (9,8)   & 170                 & $\pm1\,$                                            &4.5   & $20$       \\  \hline\hline
    \end{tabular}
    \caption{
        The transitions considered in this work along with the leading order standard model prediction  $E^{\rm{SM,LO}}_{n,n-1}\,$, the measured\,(or projected) values of $\Delta_{n,n-1}/E^{\rm{SM,LO}}_{n,n-1}$  given with\, (or without) references, the inverse decoupling radius $r^{-1}_{n_{f}}$ of the lower level in the transition, and  an estimation of the relative nuclear polarization contribution.
        } 
    \label{tab:transitions}
\end{table*}

\mysec{Contributions of nuclear polarization and new physics}.
We consider transitions between circular states in nuclei bound to negatively charged hadrons.
To increase NP sensitivity, $n$ is chosen to be as small as possible but large enough so that the effects of QCD are negligible at the relevant accuracy level, given in Table~\ref{tab:transitions}; see also~\cite{1997-Strong, Paul:2020cnx}.

Consequently, we parameterize the theoretical prediction of the $n$ level as 
\begin{align}
    \label{eq: energy contributions}
    E_{n}^{{\rm th}} 
    =
    \underbrace{E_{n}^{\rm SM-NPol}+E_{n}^{\rm NPol}}_{E_n^{\rm SM}}+E_{n}^{X},
\end{align}
where $E_{n}^{{\rm SM-NPol}}$ is the contribution of the SM excluding the NPol, $E_{n}^{{\rm NPol}}$ is the sum of the contributions of the polarizations of the nucleus and the orbiting hadron, and $E_{n}^{X}$ is the contribution of NP due to the exchange of an $X$ boson.
Hereafter, wherever $n$ is written alone, the angular momentum number is implied to be $l=n-1$.

Due to the separation of the NPol effect, we assume that the uncertainty in $E_{n}^{\rm SM-NPol}$ is either already negligible compared to the precision goals of ongoing experiments~\cite{Paul:2020cnx}, or that it would be reduced hand in hand with the experimental improvements~\cite{2025-PAX}.

For the states considered, the orbiting particle velocity is roughly $Z\alpha/n\ll1$, so the leading-order contribution to the energy levels, given in Table~\ref{tab:transitions}, is of the Coulomb type $E^{\rm SM,LO}_n=-(Z\alpha/n)^2 \mu/2$, where $\mu$ is the reduced mass. 
In the static (Born-Oppenheimer) approximation, the energy shift associated with the dipole contribution of the electric polarizability is~\cite{Ericson:1972nhh, Zatorski:2022sze}
\begin{align}
    \label{eq:Pol1}
     E_{n}^{\rm NPol}
     =
     -\frac{1}{2}\alpha
     \left(\alpha_{E}^{N}+Z^{2}\alpha_{E}^{H}\right)
     \left\langle r^{-4}\right\rangle _{n}
     \equiv
     \alpha_{E}^{{\rm tot}}h_{n}^{\rm NPol}\,,
\end{align}
where $\alpha_{E}^{\rm tot}\equiv\alpha_{E}^{N}+Z^{2}\alpha_{E}^{{H}}$ is the total polarizability, comprising that of the nucleus $\alpha_{E}^{N}$, and the orbiting hadron $\alpha_{E}^{H}$.
For $H=\left\{ \bar{p},\pi^-,K^- \right\}$, $\alpha_{E}^{H}=\left(12.0\pm0.5,2.0\pm0.9,<200\right) \times 10^{-4}\, \mathrm{fm}^{3}$~\cite{1973-KaonPol, Pasquini:2019nnx,Moinester:2022tba}, respectively.
$\left\langle r^{-4}\right\rangle _{n}$ is the expectation value of $r^{-4}$ for a state with principal quantum number $n$.
For circular states, it returns $h_{n}^{\rm NPol}=-8\mu^{4}Z^{4}\alpha^{5}\left(2n-4\right)!/(n^{4}\left(2n\right)!)$. 

To assess the order of magnitude of $\alpha_{E}^{N}$, we adopt the results of a recent global analysis of photoabsorption cross-section measurements, which returns~\cite{Orce:2023qtm}
\begin{align}
    \label{eq:polarizability}
    \alpha_{E}^{N}
    \approx 
    8\frac{\left(\frac{A}{132}\right)^{2}}{\left(\frac{A}{132}\right)^{\frac{1}{3}}-0.31}{\rm fm}^{3}\,,
\end{align}
where $A$ is the mass number of the nucleus.
Eqs.~\eqref{eq:Pol1} and~\eqref{eq:polarizability} allow us to approximate the NPol corrections listed in Table~\ref{tab:transitions}, revealing that they are not negligible at the accuracy goals of the upcoming experiments and that the contribution of $\alpha_{E}^{N}$ outweighs that of $Z^2\alpha^H_E$.
 
The uncertainties associated with these medium-mass and heavy nuclei are projected to be considerable~\cite{2017-alphad,2023-40Ca,Orce:2023qtm,2024-alphaD, gorchtein2025hitchhiker}, thus, NPol could become a bottleneck to NP searches in these systems.
Below, we show that by using two transitions, one can probe NP without having to calculate NPol, thus bypassing this difficulty.

The spin-independent component of new interactions manifests as an effective Yukawa potential~\cite{1935-Yuk}
\begin{align}
    \label{eq:potentials}
    V_{X}\left(\boldsymbol{r}\right)
    =
    (-1)^s\frac{g_{N}^{X}g_{H}^{X}}{4\pi}\frac{e^{-m_{X}r}}{r} \, ,
\end{align}
where $X$ denotes the new boson with spin $s$. 
$g_H^X\,(g_N^X)$ represents the coupling to the hadron$\,$(nucleus).
In a nonrelativistic approximation, the first-order perturbation of this potential to the energies of circular states is 
\begin{align}
   \label{eq:energy corr}
    E_{n}^{X}
    =
    (-1)^s\frac{g_{H}^{X}g_{N}^{X}}{4\pi}
    \frac{1}{r_n}\frac{1}{\left(1+\frac{m_{X}r_{n}}{2n}\right)^{2n}}
    \equiv g_{H}^{X}g_{N}^{X}h_{n}^{X},
\end{align}
where $r_n \equiv n^{2} /(Z\alpha\mu)$ represents the decoupling radius the distance beyond which the sensitivity to NP decreases.
Note that $r_{n}^{2}$ corresponds to $\left<r^{2}\right>_{n}$ when $n\gg 1$ Table~\ref{tab:transitions} lists inverse decoupling radii of the lower levels in the transitions considered here.
The transitions span a range of $11\,\keV$ for a Rydberg transition in $\bar{p}\,^4\text{He}$, which laser technology can access, up to $5.5\,\MeV$ for the $10\rightarrow9$ transition in $\bar{p}\,\text{Pb}$, situated in the x-ray regime. 
This highlights the importance of x-ray spectroscopy in investigating short-range interactions.

\mysec{Probing new physics}.
The strongest radiative transitions in the cascade of the exotic particle are those between two subsequent circular states~\cite{2008-Noble}, so that the transition energies are $E^{\rm exp}_{n,n-1}$, where for any quantity $Y$ we define $Y_{n,n-1}\equiv Y_n-Y_{n-1}$.

When the NPol uncertainty, $\sigma_{E_{n,n-1}^{\rm NPol}}$, is negligible compared to the experimental uncertainty, $\sigma_{E_{n,n-1}^{\rm exp}}$, we can use a single transition, preferably to the lowest $n$ not affected by hadronic interactions, to probe NP. 
Assuming an agreement between the experiment and the prediction of SM, the upper bound of the projected 95\% confidence level is 
\begin{align}
    \label{eq:DeltaXEnn1}
    | E^{X}_{n,n-1} |
    =
    | E^{\rm exp}_{n,n-1} - E^{\rm SM}_{n,n-1} |
    < 2\,\sigma_{E^{\rm exp}_{n,n-1}} \, .
\end{align}
Combining Eq.~\eqref{eq:energy corr} and Eq.~\eqref{eq:DeltaXEnn1} returns
\begin{align}
    \label{eq:gHgN1transition}
    [g_{H}^{X}g_{N}^{X}]_{\rm 1T}
    < \frac{2\,\sigma_{E^{\rm exp}_{n,n-1}}}{| h_{n,n-1}^X |}
    \xrightarrow[m_X r_n\to 0]{}
    4\pi Z\alpha R_\sigma\,,
\end{align}
where $R_\sigma\equiv\sigma_{E^{\rm exp}_{n,n-1}}/E^{\rm exp}_{n,n-1}$.
From Eq.~\eqref{eq:gHgN1transition}, we learn that the sensitivity is flat up to the mass of $m_X \sim 1/r_n$, and then quickly decouples as $m_X^{2n}$. 
We denote this case as \textit{single transition}, 1T. 

The second case is when $\sigma_{ E_{n,n-1}^{\rm exp}}<\sigma_{E_{n,n-1}^{\rm NPol}}$. 
As the uncertainty in $\alpha^{N}_{E}$ is expected to be much larger than the magnitude of the corrections to Eq.~\eqref{eq:Pol1}, we can simultaneously determine $\alpha^{N}_{E}$ and probe NP by considering two transitions measured for the same system.
We combine Eqs.~\eqref{eq: energy contributions},~\eqref{eq:Pol1} and \eqref{eq:energy corr} and obtain 
\begin{align}    
    \Delta_{n,n-1} 
    \equiv
    & E_{n,n-1}^{\rm exp}- E_{n,n-1}^{\rm SM-NPol}
    \nonumber\\
    =
    & g_{H}^{X}g_{N}^{X} h_{n,n-1}^{X}
    +\alpha_{E}^{\rm tot} h_{n,n-1}^{{\rm {NPol}}} \, ,
\end{align}
where $g_{H}^{X}g_{N}^{X}$ and $\alpha_{E}^{\rm tot}$ are the two unknowns. 
Next, assuming that $\Delta_{n,n-1}$ and $\Delta_{n',n'-1}$ were measured, we can solve for $g_{H}^{X}g_{N}^{X}$ and $\alpha_{E}^{\rm tot}$ and get
\begin{align}
    \label{eq:gHgNSol}
    [g_{H}^{X}g_{N}^{X}]_{{\rm 2T}}=&\frac{\Delta_{n,n-1}h_{n',n'-1}^{{\rm NPol}}-\Delta_{n',n'-1}h_{n,n-1}^{{\rm NPol}}}{h_{n,n-1}^{X}h_{n',n'-1}^{{\rm NPol}}-h_{n',n'-1}^{X}h_{n,n-1}^{{\rm NPol}}}\,,\\\label{e:alphaE1totSol}[\alpha_{E}^{{\rm tot}}]_{{\rm 2T}}=&\frac{\Delta_{n',n'-1}h_{n,n-1}^{X}-\Delta_{n,n-1}h_{n',n'-1}^{X}}{h_{n,n-1}^{X}h_{n',n'-1}^{{\rm NPol}}-h_{n',n'-1}^{X}h_{n,n-1}^{{\rm NPol}}}\,.
\end{align}
Thus, by measuring two transitions and combining with theory, we can determine both NPol and NP.
For every pair of transitions, there is a specific $m_X$ for which $h_{n',n'-1}^{X}h_{n,n-1}^{\rm NPol}= h_{n,n-1}^{X}h_{n',n'-1}^{\rm NPol}$, and the sensitivity is lost.
The sensitivity to $g_H^X g^X_N$ is estimated using error propagation of Eq.~\eqref{eq:gHgNSol}.
For simplicity, we assume that the uncertainties on $\Delta_{n,n-1}$ and $\Delta_{n',n'-1}$ are not correlated and are of the same fraction of the energies.
This case is denoted as \textit{dual transition}, 2T.

For heavy nuclei, a non negligible contribution from the quadrupole electric polarizability is possible~\cite{Ericson:1972nhh}.
Its contribution to the potential scales as $r^{-6}$, and the relative effect compared to the dipole polarizability is $(r_A/r_n)^2$, where $r_A \approx 1.2 A^{1/3}$~fm is the RMS nuclear charge radius~\cite{Ericson:1972nhh}. 
It is expected to be small for our chosen transitions and nuclei but can still be systematically fitted from the data using additional transitions. 
Moreover, different isotopes could be probed to better understand this effect.

\begin{figure}[t]
    \centering
    \includegraphics[
    trim=0cm 0cm 3cm 0cm,
    width=0.99\linewidth]{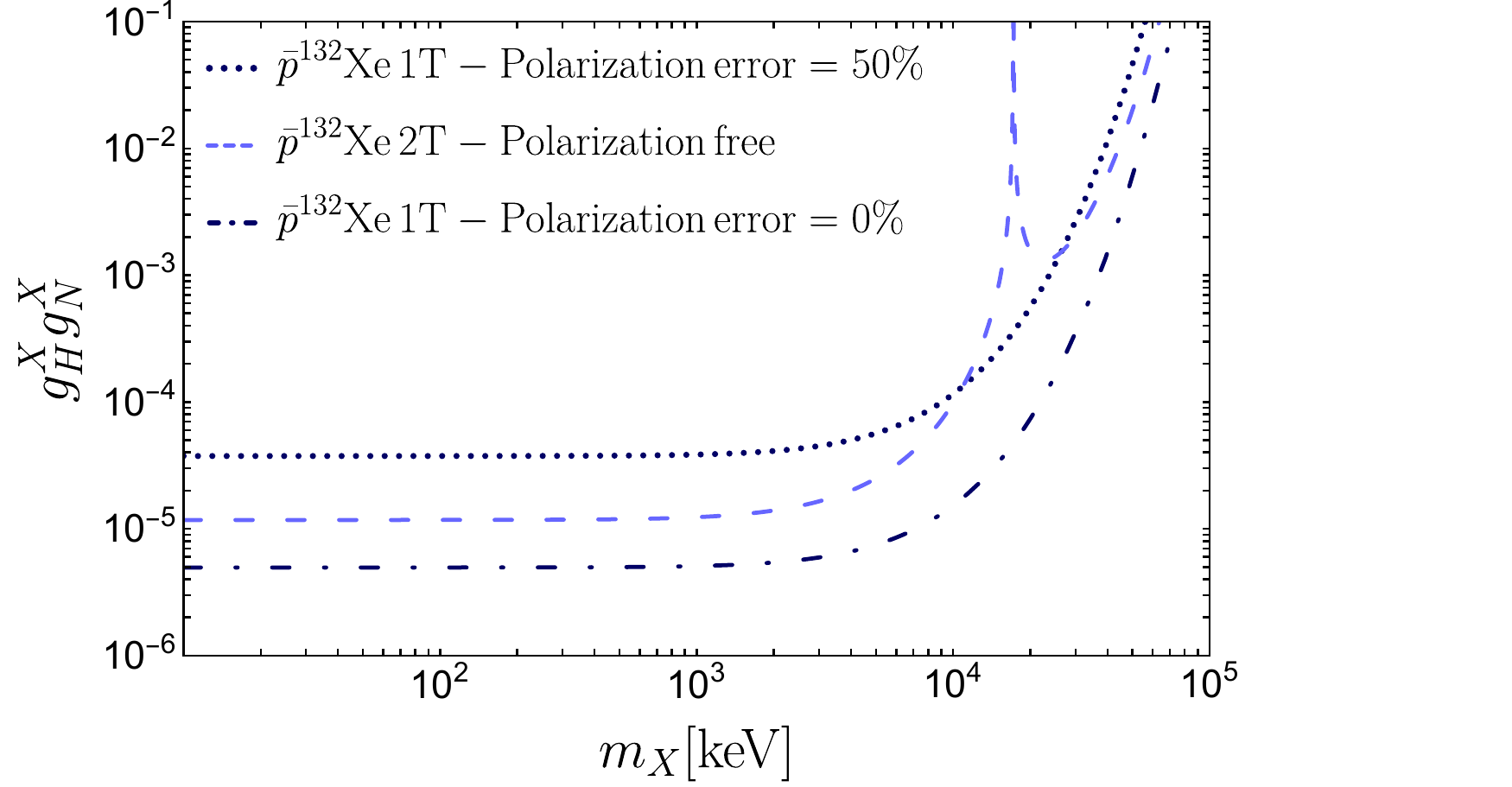}
    \caption{
    $\bar{p}^{132}\rm{Xe}$ projections with $R_\sigma=10^{-6}$. 
    The projected bound in dotted\,(dot-dashed) blue is for the $\left(10,9\right)\rightarrow\left(9,8\right)$ transition  with\,(without) 50\% polarizability error that constitutes a 10\,ppm uncertainty to the NPol  contribution.
    In dashed light blue we show the bound for combining two transitions: $\left(11,10\right)\rightarrow\left(10,9\right)$ and $\left(10,9\right)\rightarrow\left(9,8\right)$ to reduce the sensitivity to NPol.}
    \label{fig:two transition example}
\end{figure}

In Fig.~\ref{fig:two transition example}, we illustrate the use of 1T Eq.~\eqref{eq:gHgN1transition} and 2T Eq.~(\ref{eq:gHgNSol}) by plotting the projection for $\bar{p}^{132}{\rm Xe}$ assuming the experimental accuracy of $R_\sigma=1\,$ppm. 
Three cases are considered:
(i)~1T: $10\to 9$ neglecting the error on the nuclear polarizability;
(ii)~1T: $10\to 9 $ with 50\,\% error on the nuclear polarizability;
(iii)~2T: $11\to 10$ and $10\to 9$ which is free of NPol. 
Case (i) is best only when NPol is controlled to at least 11\% of its value, which is beyond current knowledge for a heavy nucleus such as $^{132}$Xe. 
However, case (iii) is the most sensitive when a realistic uncertainty is assumed, as in case (ii).

\begin{figure}[tbp!]
    \centering
    \includegraphics[ 
    trim=2.6cm 16cm 3.0cm 0cm, clip, width=0.99\linewidth]{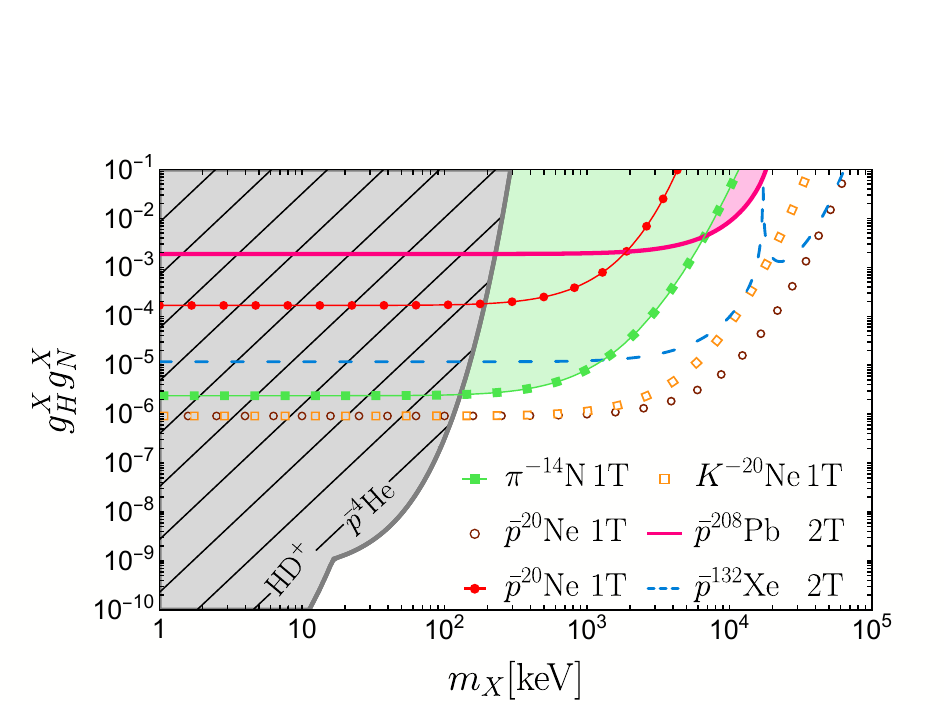}
    \caption{
    New bounds\,(shaded solid) and projections\,(dashed) from this work compared to existing bounds from $\bar{p}\,^4$He and HD$^{+}$\,(shaded gray area).} 
    \label{fig:new gHgN bound}
\end{figure}

\mysec{Current data and bounds}.
We consider the available data and ongoing experiments in hadronic atoms and derive novel bounds and future projections on NP; see Fig.~\ref{fig:new gHgN bound}.
They are given in the $m_X-g^X_H g^X_N$ plane and can be matched to specific models as shown for two examples below. 

We begin with an analysis of mesonic atoms, which have so far been used to extract meson masses~\cite{JECKELMANN1986709, 1994-PiMASS, 1996-PionMaSS, LENZ199850, Trassinelli:2016kki, 1973-KaonPol} and measure strong interaction effects~\cite{1985-StrongPi, 1998-StrongPi, 1997-Strong, hirtl2021redetermination, 2022-KHe}.
To utilize them for NP searches, we adopt the pion pass $[m_{\pi}]_{\rm kin}=139.57021(14)\,\MeV$, accurate to $1.0\,$ppm, from kinematic measurements~\cite{1994-PiMASS, 1996-PionMaSS,2021-PiPlus}.
The most precise single transition measurement in pionic atoms is of the $5\to 4$ line in $\pi^-{}^{14}$N and returns $[m_{\pi}]_{5\to4}=139.57077(18)\,\MeV$\,\cite{Trassinelli:2016kki}, accurate to $1.3\,$ppm.
Because the transition energy is proportional to $m_\pi$ at the percent level, we deduce $\Delta_{5,4}/E^{\rm{SM,LO}}_{5,4}\,=3.9\pm1.7\,$ppm. 
Other high-precision measurements in pionic and kaonic atoms produce weaker bounds~\cite{1973-KaonPol}, are affected by residual electrons~\cite{JECKELMANN1986709, Daum:2021wol, 2005-QDEK}, or by QCD~\cite{1985-StrongPi, 1998-StrongPi, hirtl2021redetermination, BIRD1983482, PhysRevC.40.2154, 2022-KHe}.

The SIDDHARTA-2 experiment at the DA$\Phi$NE collider recently demonstrated high-resolution spectroscopy of circular transitions in $K^-\,$Ne~\cite{2024-KNe}, of which the one with the highest energy is $6\rightarrow5$.
For a future projection of the sensitivity of this transition, we consider a $1\,$ppm accuracy and also assume that the kaon mass is measured independently, e.g. by laser spectroscopy~\cite{mklaser}. 

Next, we consider transitions in antiprotonic atoms.
In Ref.~\cite{1999-StronPBAR}, the strong interaction shift of the 2p manifold in $\bar{p}\,$D was measured in two ways: 
(i)~Comparing with the known $k\alpha_1$ x-ray fluorescence line in atomic sulfur. 
(ii)~Comparing with the $\bar{p}\,^{20}$Ne $13\rightarrow12$ line whose energy was calculated assuming that there is no NP.
As both methods return the same $\bar{p}\,$D strong interaction shift, they can be reinterpreted as a measurement of $\bar{p}\,^{20}$Ne $13\rightarrow12$ energy calibrated by the sulfur line.
From the information given in Table~9 of Ref.~\cite{1999-StronPBAR} we deduce $\Delta_{13,12}/E^{\rm{SM,LO}}_{13,12}=14\pm23\,$ppm given in Table~\ref{tab:transitions}.

To probe higher new boson masses, we consider the transitions measured in $\bar{p}$\,Pb~\cite{1975-PbExp}, where a natural abundance sample was used.
Since the lines for different isotopes were not experimentally resolved, the NPol may be treated as an average over them.
A calculation of the corresponding energies is given in~\cite{Borie:1983nlf}. 
However, it does not include the effect of nuclear polarization.
A rough estimation for $^{208}$Pb$(11,10)\rightarrow(10,9)$ using Eq.~\eqref{eq:Pol1} returns $\sim40\,$ppm.
As this is of similar magnitude as the experimental error, we remove it employing the 2T method to search for NP, considering the two lowest-lying measured intervals $11\to10$ and $10\to9$.
An agreement is found between experiment~\cite{1975-PbExp} and theory~\cite{Borie:1983nlf} resulting in the bound plotted in Fig.~\ref{fig:new gHgN bound}.
Although the accuracy is not as high as in lighter systems, the compact nature of this system allows one to competitively probe a new boson mass of $1/r_n\approx6\,\MeV$.
Other high-precision measurements in antiprotonic atoms either produce weaker bounds (\cite{2003-PbarNe} and Table~IX in~\cite{1977-StrongMAss}), or are affected by QCD~\cite{Simons_1988, AUGSBURGER1999149, ANAGNOSTOPOULOS1999c305, KOHLER1986327, POTH1987667,1977-StrongMAss}.

Having established that x-ray spectroscopy in hadronic atoms is sensitive to the $0.1-10\,\MeV$ NP scale, we consider the up-and-coming PAX experiment at the CERN antiproton decelerator~\cite{2024-Prospects,2024-169,2024-PAX,2025-PAX}.
Suitable transitions, where QCD contact terms and finite-size effects are negligible, were identified in Ref.~\cite{Paul:2020cnx} in the context of the probing of high-field QED.
Considering the existing bounds discussed above, an accuracy better than $100\,$ppm would already probe a new parameter space.
To show the full potential of these measurements, we consider an accuracy of $1\,$ppm for two representative cases: 
(i)~1T for $6\rightarrow5$ line in $\bar{p}^{20}$Ne. 
Since $E^{\rm NPol}_{6,5}\approx 1\,{\rm ppm}$, it is not expected to be the bottleneck; 
(ii)~2T combining $12\rightarrow11$ and $11\rightarrow10$ in $\bar{p}^{132}$Xe, as $E^{\rm NPol}_{11,10}\approx20\,{\rm ppm}$ it can potentially limit the sensitivity of a single-transition measurement.

The existing strongest bound from exotic atoms is from $\bar{p}\,^4$He~\cite{ASACUSA:2016xeq,Germann:2021koc}, which decouples at $20\,\keV$, as shown in Fig.~\ref{fig:new gHgN bound}.
Other probes of new hadronic forces are from molecular spectroscopy ($m_X\lesssim10\,\keV$)~\cite{2014-Karr,SchillerHDp} and neutron scattering ($m_X\lesssim100\,\keV$)~\cite{Nesvizhevsky:2007by}. 
There are stringent stellar cooling limits~\cite{Hardy:2016kme,Bottaro:2023gep} that decouple at $100\,\keV$ and from supernova 1987A bounds~\cite{Hardy:2024gwy} that reach higher masses and relevant to smaller couplings. 
However, these are  model dependent and subject to different systematics, see \eg~\cite{OHare:2020wah,Burrage:2016bwy,Masso:2005ym,Jaeckel:2006xm,DeRocco:2020xdt,Budnik:2020nwz,Bloch:2020uzh}.
Our updated bounds and projections reach a greater mass than previous probes, up to $\sim10\,\MeV$.

\begin{figure*}[t]
    \centering
    \includegraphics[trim=3.4cm 16cm 2.2cm 0cm, clip, width=0.49\linewidth]{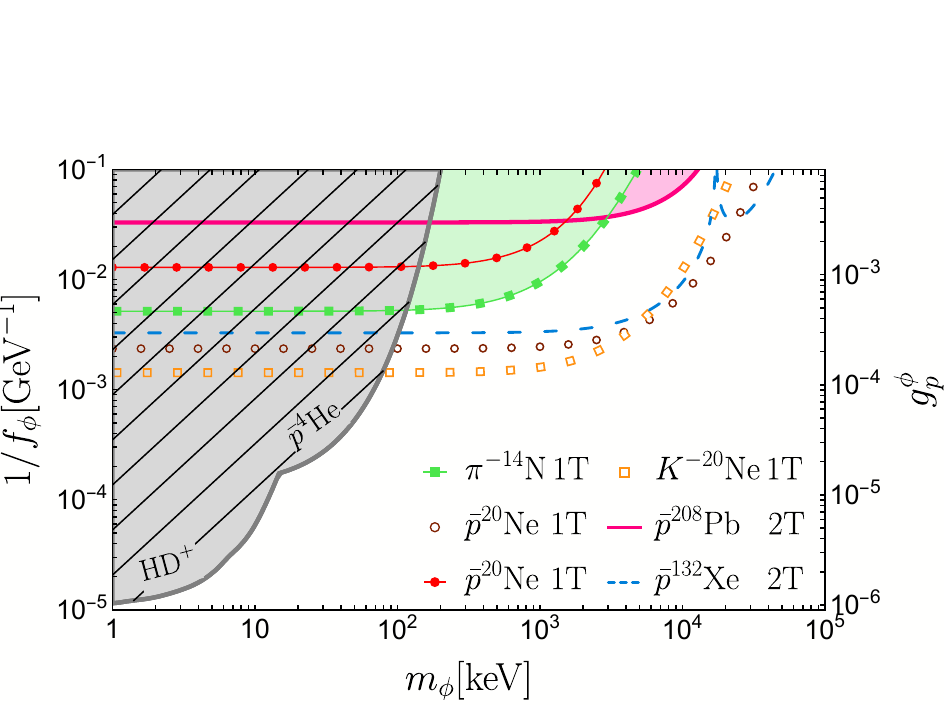}
    \includegraphics[trim=2.6cm 16cm 3.0cm 0cm, clip, width=0.49\linewidth]{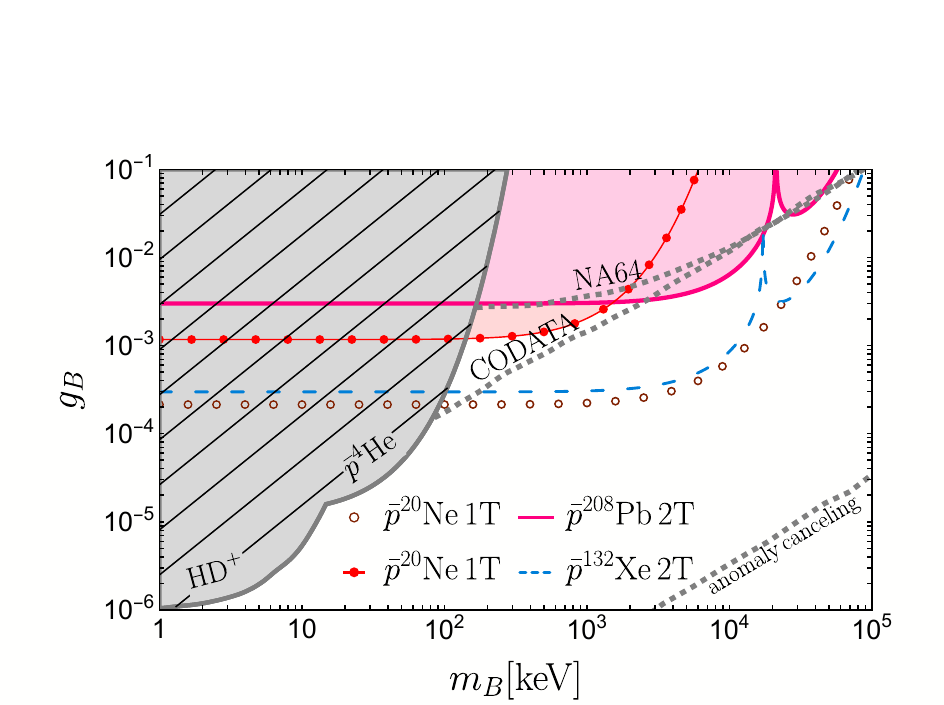}
    \caption{
    Same as Fig.~\ref{fig:new gHgN bound} with  
    Left: the $uds$-scalar model. 
    Right: the $B$-boson model with additional
    bounds from NA64, CODATA, and new anomaly-canceling fermions~(see main text for details).}
\label{fig:BenchmarkModels}
\end{figure*}

\mysec{Application to benchmark models}.
The above bounds can be cast into specific BSM models; here are the results for two benchmark models.
The first is the $uds$-scalar from Ref.~\cite{Delaunay:2025lhl} and the second is the $B$-model (where $B$ stands for a gauged baryon number) assuming predominantly invisible decays.
In the following, we show that exotic atoms can be the leading probes of these models for a mediator mass in the $100\,\keV$ to $10\,\MeV$ range. 
Note that the Higgs-mixing model is severely constrained by $K^+\to\pi^++\textrm{inv.}$~\cite{NA62:2021zjw,Goudzovski:2022vbt}.

In the $uds$-scalar model, a new CP-even scalar, $\phi$, with mass $m_\phi$, exclusively interacts with the $u$, $d$, and $s$ quarks and decays invisibly.
The couplings are proportional to the quark masses and are suppressed by the cutoff scale $f_\phi$.  
As shown in Ref.~\cite{Delaunay:2025lhl}, the stringent bounds from $K\to\pi \phi$~\cite{NA62:2021zjw,KOTO:2024zbl} are evaded to leading-order in the chiral expansion and thus relaxed, unlike in the case of scalar Higgs mixing. 
The effective $\phi$-quarks interaction is given by
\begin{align}
    \label{eq:Luds}
    \cL^{\rm int}_{\phi} 
    = 
    \frac{\phi}{f_{\phi}}\sum_{q=u,d,s}m_q \bar{q} q \, ,
\end{align}
without coupling to other SM particles. 
Following \eg~\cite{Bishara:2017pfq,Batell:2018fqo}, we match $\cL^{\rm int}_\phi$ to the hadron level 
\begin{align}
    \cL_{\phi}^{\rm int,had}
    \subset
    \phi\big(
    &g_{p}^{\phi}\bar{p}p+g_{n}^{\phi}\bar{n}n  \nonumber\\
    &+2m_{\pi}g_{\pi}^{\phi}\pi^{+}\pi^{-}+2m_{K}g_{K}^{\phi}K^{+}K^{-}\big),
\end{align}
with $g^\phi_{\pi(K)}=-\frac{m_{\pi(K)}}{2f_\phi}$, $g^\phi_{p(n)}=\frac{m_{p(n)}}{f_\phi}\sum_qf^{p(n)}_q$ and $f^{p(n)}_u=0.018(0.016)$, $f^{p(n)}_d=0.034(0.038)$, and $f^{p(n)}_s=0.044(0.044)$~\cite{Bishara:2017pfq}. 
The interaction with nuclei is given by $g_{N}^{X} \approx (A-Z)g_{n}^{X}+Zg_{p}^{X}$.

The bounds on the $uds$-model are plotted in the left panel of Fig.~\ref{fig:BenchmarkModels} in the $m_\phi-f^{-1}_\phi\,(g_{p}^{\phi})$ plane, where we projected the bounds of Fig.~\ref{fig:new gHgN bound}.
The approximated NLO bounds from rare kaon decay are $1/f_\phi \lesssim0.8/\GeV$~\cite{Delaunay:2025lhl} and are not shown. 
We note that the probed range of parameter space is mildly fine-tuned, since for a TeV cutoff, the 1-loop correction to $m_\pi$ is $\delta m_\phi^2 \sim \TeV^2 (m_s/f_\phi)^2/16\pi^2$. 
We learn that heavy exotic atoms already set the strongest bounds on the $uds$ model for $0.1 \lesssim m_\phi\lesssim10\,\MeV$ with the potential to improve by at least an order of magnitude. 

The $B$-boson model introduces a new massive spin-1 gauge boson, $B$, with universal vector-like couplings to the quarks.
The SM charged leptons are coupled via 1-loop kinetic mixing and we assume $\BR(B\to {\rm inv.})\approx1$. 
This model was analyzed in several places, \eg~\cite{Ilten:2018crw,Tulin:2014tya}. 
The interactions with the SM fermions are given by
\begin{align}
    \label{eq:LB}
    \cL_B^{\rm int} 
    = 
    \frac{g_B}{3} B^\mu \sum_{q}\bar{q}\gamma_\mu q 
    -\frac{g_B e^{2}}{48\pi^2} B^\mu\sum_\ell\bar{\ell}\gamma_\mu \ell \,  , 
\end{align}
where $q\,(\ell)$ is the SM quarks\,(charged leptons).
The $\cL_B^{\rm int}$ can be mapped to the hadron interactions 
\begin{align}
    \cL_{B}^{\rm int,had}
    =
    g_B B_\mu\left( \bar{p}\gamma^\mu p + \bar{n}\gamma^\mu n \right)\,.
\end{align}
The baryon number is an anomalous symmetry, and thus requires additional degrees of freedom.
Such considerations lead to strong bounds from rare bottom and kaon decays~\cite{Dror:2017ehi,Dror:2017nsg,Dobrescu:2014fca}. 
However, a similar model that is anomaly-free can be built by mixing the SM fermions with heavy vector-like fermions as in \eg~\cite{Kamenik:2017tnu,Delaunay:2020vdb}. 

We plot our new $\bar{p}\,$Ne and $\bar{p}\,$Pb bounds and the projections for the $B$ model in the right panel of Fig.~\ref{fig:BenchmarkModels}.
We compare our result to the existing bounds from $\bar{p}^4$He, NA64~\cite{NA64:2022yly} and to rescaling of the $B-L$ bound from~\cite{Delaunay:2022grr}, denoted as CODATA.
The last two are based on a loop-induced $B$-electron coupling that depends on the model's UV details and can be adjusted accordingly. 
We learn that $\bar{p}$Ne and $\bar{p}$Pb are already the robust leading probes of this model and that future measurements have the potential to increase sensitivity by an order of magnitude.

\mysec{Conclusions}.
We explored the potential of precision spectroscopy of heavy exotic atoms in which electrons are replaced by $\bar{p}$, $\pi^-$ or $K^-$, to probe new force carriers with hadronic couplings.
We have considered transitions that are clean from short-range hadronic effects, thus, the SM contribution can be predicted to high accuracy.
However, nuclear polarizability can still have sizable effects on these transition lines.
Therefore, we propose to use two lines and simultaneously solve for new-physics and polarizability effects. 
Based on the available data in $\bar{p}\,$Ne, $\bar{p}\,$Pb and $\pi^-\,$N, we extract new world-leading bounds on two benchmark models and derive projections for future measurement for several representative systems.
For mediator masses of up to $10\,\MeV$, future experiments have the potential to improve the sensitivity by up to two orders of magnitude.
In addition, while this work considered only spin-independent interactions, spin-dependent interactions can be probed in a future work. 

\begin{acknowledgments}
We thank N.~Paul, K.~Pachucki, M.~Gorshteyn, and C.~Delaunay for illuminating discussions and comments on the manuscript.
Special thanks goes to P.~Indelicato for valuable input on the existing experimental data and for testing prior calculations for consistency.
The work of H.L. is supported by the U.S. Department of Energy under Grant Contract DE-SC0012704. 
The work of B.O. is supported by the Helen-Diller quantum center and ISF (grant no. 2071390).
The work of O.S. and Y.S. is supported by grants from the NSF-BSF (grant No. 2021800), the ISF (grant No. 597/24).
The work of O.S. is also supported by the Helen-Diller quantum center.
\end{acknowledgments}

\bibliographystyle{utphys28mod}
\bibliography{ref.bib}

\end{document}